# When Life Gives You Lemons, Squeeze Your Way Through: Understanding Citrus Avoidance Behaviour by Free-Ranging Dogs in India

Tuhin Subhra Pal [1], Srijaya Nandi [1], Rohan Sarkar [1], Anindita Bhadra [1*]

[1] Department of Biological Sciences, Indian Institute of Science Education and Research, Kolkata

*Corresponding Author

Anindita Bhadra (abhadra@iiserkol.ac.in)

**Abstract**

Palatability of food is driven by multiple factors like taste, smell, texture, freshness, etc. and can be very variable across species. While foraging, animals use different mechanisms to locate their preferred food, and local factors can sometimes determine food habits of different populations. There are classic examples of local adaptations leading to speciation, driven by food availability. Urbanization across the world is leading to rapid decline of biodiversity, while also driving local adaptations in some species. Free-ranging dogs are an interesting example of adaptation to a human-dominated environment across varied habitats. They have co-existed with humans for centuries, and present a perfect model system for studying local adaptations. In this study, we attempted to understand a specific aspect of their scavenging behaviour in India – citrus aversion. Pet dogs are known to avoid citrus fruits and food contaminated by them. In India, lemons are used widely in the cuisine, and also discarded in the garbage with leftovers from the kitchen. Hence, free-ranging dogs, that typically are scavengers of human leftovers, are likely to encounter lemons and lemon-contaminated food on a regular basis. We carried out a population level experiment to test free-ranging dogs' response to chicken contaminated with various parts of lemon. The dogs avoided chicken contaminated with lemon juice the most. Further, when provided with chicken dipped in three different concentrations of lemon juice, the lowest concentration was most preferred. The dogs used different manoeuvres to procure the chicken and avoid the lemon. A survey conducted at the field sites revealed that the local people use lemon in their regular diet extensively and also discard these with the leftovers, where the free-ranging dogs' forage. People were found to avoid giving citrus contaminated food to their own pets, but did not follow the same caution in case of free-ranging dogs. This study revealed that free-ranging dogs in West Bengal, India, are well adapted to scavenging among citrus-contaminated garbage, and have their own strategies to avoid the contamination as far as possible, while maximizing their preferred food intake. These results contribute to a better understanding of FRDs' dietary preferences and inform strategies for promoting responsible human-animal interactions by offering insightful information about their foraging behaviours.

**Introduction**

Food is the primary requirement for the survival of any organism. While some species are generalists in their feeding habits, others are specialists, with very specific food preferences. Feeding habits can lead to specific adaptations, which may eventually lead to speciation, as in the case of the finches of the Galapagos islands (Losos & Ricklefs, 2009). While food preference can be shaped by evolutionary processes, the decision to eat a specific food item can be driven by local parameters, like taste, flavour and visual appeal (Cole & Endler, 2015; Provenza, 1995). These factors can determine the palatability of a food item and make it more or less appealing to an individual.

The flavour of food is a sensory property that is mostly determined by taste and smell, and it influences the food's hedonic value or palatability. It significantly impacts the type and quantity of food consumed, also known as appetence (Bellisle, 1989; Baumont, 1996; Ishii et al., 2003). When given choices, animals show their food preferences through intake ratios or consuming behaviour, in contrast to, humans who can communicate their preferences verbally. Therefore, diet selection is a means to measure animals' food preferences (Forbes & Kyriazakis, 1995).

Palatability is influenced by both taste order and surface characteristics of its components (Kenney & Black, 1984). In addition, palatability can be altered through dietary exposure (Chapple and Lynch, 1986; Provenza & Balph, 1987). Habitat alterations can lead to changes in the availability of preferred food, and thus can impact the survival of species. During foraging, animals make assessments about what to eat, where to find it, how much time to spend in different places, and how to move between patches (Stephens, 2008). Classical optimal foraging models (Pyke et al., 1977) indicate that varying food availability in nature impacts these decisions and requires animals to adapt and adjust in unanticipated conditions (Charnov & Orians, 2006). In the recent times, urbanization and habitat loss have exposed many species of animals across taxa to altered environments. While this often leads to loss of species diversity, some animals adapt remarkably to their new habitats, and studying these animals can provide insights into the mechanism of adaptation to altered habitats. Free-ranging dogs (FRDs) that occupy all possible habitats around human settlements across the Global South (Biswas et al., 2023), are an excellent model system for understanding urban adaptation. Since they are primarily scavengers, dependent on human-generated waste, the food available to them is influenced by the diet of the human community sharing the habitat with them. Such food can vary in their appeal to the FRDs, due to varied reasons. The domestic dogs share a common ancestry with grey wolves, which are hunters. The shift in behaviour of the FRDs to a scavenging habit, and the ability to digest carbohydrates is associated with the domestication process (Axelsson et al., 2013; Bhadra & Bhadra, 2014)

FRDs possess exceptional scavenging skills and are flexible in their food choices, varying from vegetable leftovers to meat (Bhadra et al., 2016). They set aside their preference for higher quality food, when confronted with high and low-quality options, probably because of familiarity (Vicars et al., 2014; Cameron et al., 2021) , or to avoid competition. Their ability to scavenge efficiently arises from a simple 'Rule of Thumb' they use to locate protein-rich food using their exceptional sense of smell (Bhadra et al., 2016; R. Sarkar et al., 2019).

Olfactory cues are crucial for decision-making in dogs. They are capable of using their noses to selectively find and eat their preferred food from garbage (R. Sarkar et al., 2019). They can discriminate between varying quantities of food in choice tasks (Banerjee & Bhadra, 2019).

Animals such as cows, lambs, and goats can learn to avoid plants or foods that cause illness (Zahorik et al., 1990). Their foraging behaviour is influenced by both internal and external factors, and is an adaptive reaction to the demands of their internal environment as well as the influence of the external environment (Blundell et al., 1985; Johnson & Collier, 2001; Melvin P. Enns, 1983) For example, rats as well as primates, prefer sweet-tasting food, which usually imply high calorific content, while avoiding bitter or sour foods, which may refer to toxicity, unripens, or spoilage (Mela, 1999). Taste aversion has been documented in varied species like rats (Nachman & Ashe, 1973) some carnivores (Lunceford & Kubanek, 2015; Nicolaus & Nellis, 1987) cattle (Olsen et al., 1989) and sheep (Thorhallsdottir et al., 1987).

In India, there is a large population of FRDs living in a variety of human habitats (Sen Majumder et al., 2016). These canines often consume a carbohydrate-rich, omnivorous diet, which mirrors local human dietary behaviours(Bhadra et al., 2016b). They are very adept at sniffing out any food smelling of meat and show a strong preference for meat even while scavenging from garbage (R. Sarkar et al., 2019). Lemons are a common component of the Indian diet (Yasmeen, 2019; P Sarkar et al., 2015; Barbosa-Cánovas et al., 2019) hence they are frequently discarded in garbage bins. Scavenging through waste and discarded food continues to be the primary means of sustenance for FRDs (Hart, 2023). As a result, lemons are frequently encountered while scavenging, which can lead to confusion when paired with other food items. Due to the ubiquitous presence of lemons in the human diet in India, we expect much of the food available to FRDs in garbage bins to be contaminated by lemon juice, pulp or both. Pet dogs, who rely on their humans for food (Bhattacharjee et al., 2017), rarely experience lemons in their natural state and tend to display apathy towards them (Woodford & Griffith, 2012). There is no information on whether FRDs reject food that is laced with or contaminated by lemon or lemon juice. Given the high competition that the FRDs face, not only from conspecifics, but other scavengers, and the irregular nature of the food available to them, it would be mal-adaptive for FRDs to reject high quality food like chicken, if it is contaminated by lemon. However, they might have an evolutionary constraint to avoid such food.

This study aimed to assess FRDs' aversion, if any, to citric food; and further, whether their avoidance behaviour varied according to the components and concentrations of lemon. We hypothesized that the dogs would avoid food contaminated by lemon, even at the cost of losing a rich source of nutrition. In addition to carrying out controlled choice tests on FRDs, we surveyed for the opinion of the human community on providing food mixed with citrus substances to dogs, including whether it was offered and the reasons behind their decisions.

## Methods

### Study Sites

In this study, we focused on adult FRDs as our subjects. The experiment took place in 17 different locations within the Nadia district (22.9747° N, 88.4337° E) of West Bengal (Supplementary Fig 1). The experiments were conducted in two time slots: 06:00 -12:00h and 15:00 - 20:00h. All tests were conducted on dogs that appeared to be healthy, i.e., did not have any obvious signs of disease or injury, and those that participated in the experiment by their choice. Human perception surveys were conducted in the same areas.

**Study subjects**

The study was carried out on adult FRDs that appeared to be healthy, i.e., showed no obvious signs of disease or skin ailments, and were not pregnant or lactating. The sex of each dog was recorded during the experiment. The survey was conducted on adult humans in the age group of 18 to 75 years, who willingly participated in the survey.

**Human perception survey**

A survey was conducted in Nadia, West Bengal, India, to gather information on people's perception of FRDs and their diet, in the areas where the experiments were conducted. The survey was conducted across ten locations, between 05:30h and 21:00h on various days. The participants included both pet owners and non-owners. An online Google© Form (Supplementary information 1) with objective-type questions was used for the survey, and the responses were recorded in the form by the experimenter. The target sample size was 500, and participants could withdraw voluntarily. Each respondent was required to answer at least 19 questions for successful completion of the survey.

**Choice tests with free-ranging dogs**

**Familiarization Phase**

This was common across all experiments described below. The familiarization test (Bhattacharjee et al., 2020) involved offering one fourth of a Marie biscuit to a FRDs. If the dog showed an interest in the biscuit, it was used as a subject for the choice test. For all the choice tests, we isolated the focal dog (the one which responded in the familiarization phase first) from its group by creating a physical barrier, which usually involved another person blocking the other dogs from approaching the set-up. Also, during the experiment, to mitigate biases in all experiments, we consciously avoided making eye contact with the dogs and refrained from pointing out anything.

1. **Two-bowl choice test**

Fresh lemon or lemon juice, Marie biscuits, and paper plates were the materials used across all the experiments. One-fourth of a Marie biscuit and one-fourth piece of a lemon were placed on two paper plates (Diameter-12cm, Height-3cm, Volumn-200ml, Biodegradable) for the trial. This ensured that the two pieces offered were nearly similar in size, when seen from a distance. The experimenter placed the two plates containing biscuit and lemon before the dog, maintaining a distance of approximately 0.5 meters between the plates. The experimenter stepped back (~1 meter) to allow the dog to approach the set-up. Three consecutive trials were conducted for each dog, providing two types of foods on different plates and placing them on two sides in random order. The individual dogs were randomly selected for testing, depending on availability and accessibility. The experiment was recorded using a mobile phone camera by another experimenter, capturing the plates, the dog, and the experimenter. Biases were minimized by avoiding eye contact, refraining from petting, isolating specific dogs, and preventing public disturbance during the experiment, ensuring the integrity of the dog's food choices. The study was carried out on 70 adult FRDs.

2. **Three-bowl lemon parts choice test**

This experiment aimed to investigate the response of FRDs to three different parts of lemon - juice, rind, and pulp, using a three-choice test format (Araujo et al., 2004, Bhadra and Bhadra 2015). Each of the three paper plates contained 10gm of solid chicken with 20ml of lemon juice (LJ), lemon pulp (LP) or lemon rind (LR), offering discrete combinations to the adult FRDs. The plates were set on a piece of cardboard (0.65 m length, 0.1 m width), spaced 0.3m apart, and placed in a random order. Dogs were given 1 minute to approach the setup. After their first sniff to the set-up, the behaviours of the dogs were video recorded for 90 seconds. The experiment was conducted on 100 adult FRDs different from the earlier set. A mobile camera was used to record videos of the bowls, the dog, and the experimenter.

3. **Three bowl, three concentration choice test**

Preparation LJ solution: Fresh lemon juice was extracted and diluted using distilled water to create different solutions such as 50%, 33.3% and 25% volume by volume, labelled 1, 2, and 3 respectively. We measured the pH of these solutions using a pH meter, to determine how acidic they were.

Trials were conducted using a setup with three bowls spaced equally on a cardboard piece as before. The bowls contained lemon juice 1, 2, and 3 in which a fresh chicken piece weighing 15 grams was placed. The order of placing bowls 1, 2, and 3 were randomized for each trial. 126 dogs from 14 locations were tested and for each trial; after placing the setup (positioned about 1-1.5 meters in front of each dog), we observed the dogs for 180 seconds. Precautions such as random dog selection and minimizing disturbances were implemented to reduce biases. After each trial, the bowls were replaced.

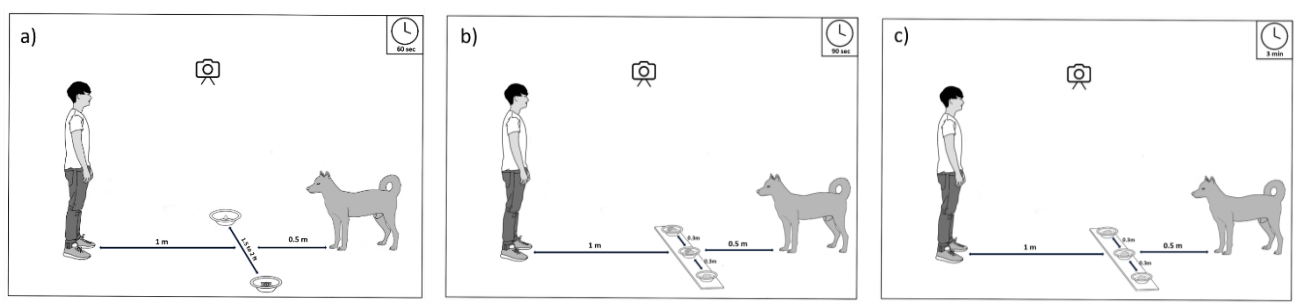

Fig. 1 Schematic representation of the experimental set-up. a) Two-bowl choice test between biscuit and lemon b) Three-bowl lemon parts choice test among chicken with LP, LR and LJ c) Three-bowl choice test among chicken with 25%. 33.3% and 50% lemon juice solution.

**Behavioural analysis**

The videos for the three-bowl three concentrations choice test were coded by three decoders, while the remaining two experiments were coded by TSP alone. Latency, defined as the time between placing the bowl or setup and the dog's initial response with a first sniff, was a key metric during the decoding process. The various behaviours used to code the responses of the dogs in these experiments have been provided in the ethogram (Table 1). Typically, the dogs sniffed the food, and chose to inspect another option or either lick or eat the food sniffed.

Sometimes the dogs refrained from immediately eating a food item after sniffing or licking it, instead manipulating it with various body parts. Such manipulation of the food was demarcated as "strategizing" The Sniff-&-Snatch (SnS) strategy is defined by eating a piece of food immediately after sniffing eat, without any intermediate behaviours. This was also seen in some cases.

Individual FRDs' responses to the different choices presented were coded. Sniffing events were labelled as s1, s2, and s3 for first, second, and third sniffs, with codes such as Sn50, Sn33.3, and Sn25 indicating sniffing occurrences at the 50%, 33.3%, and 25% lemon juice concentrations, in case of the three-bowl three concentrations choice test. Licking events likewise were represented by Lc50, Lc33.3, and Lc25; strategizing by St50, St33.3, and St25; and eating by ET50, Et33.3, and Et25. Behavioural instances for each dog were tallied from 1 to 12 for all the three choices taken together. The total duration of each dog's interaction with the contents of each bowl, from the first sniff to the end of eating, was recorded in seconds. Furthermore, the time taken for the first engagement, commencing with the initial sniff and concluding when the dog moved its mouth more than ~ 0.07 m away from the bowl or lost interest, was measured in seconds.

| Behaviour | Definition |
|---|---|
| Sniffing | The dog brings its nose within 0.1 m of the bowl to investigate the scent of the contents. |
| Licking & Vigorous licking | Involves the active use of the tongue to taste or clean the food or liquid present in the bowl. Vigorous licking refers to instances where dogs lick more than 3-4 times consecutively. |
| Strategizing | Observed when dogs purposefully refrained from immediately consuming a food item after sniffing or licking it, instead manipulated it with various body parts like tongue, foreleg, muzzle etc. |
| Eating | Actively taking food into the mouth, resulting in complete absorption or swallowing. |

Table. 1 Ethogram: A list of behaviours coded from experiment videos and their definitions.

**Statistical analysis:**

We performed Chi-Square test with a defined significance criteria of $p < 0.05$ to detect differences between various occurrences like sniffing, licking etc. To assess the time data, which was recorded in seconds, we used the Kruskal-Wallis's test. Dunn's test was used to carry out post-hoc pairwise comparisons following the Kruskal-Wallis's test. We divided the time investment for each of the three bowls by the total length of time investment to convert the time investment data into proportional data, which had originally been recorded in seconds.

We used beta regression using the 'betareg' package to analyse the influence of concentration, latency, eating events and gender on the first interaction time and the total time investment. We also performed a logistic regression to analyse the influence of concentration, latency, gender, the first interaction time and total time investment on the likelihood of eating events. Performance package was used to analyse the model's performance. We used the statistical software R Statistical Software (v4.2.2; R core team 2022) to perform statistical tests and models.

**Results**

### A. Understanding Perceptions and Practices in Feeding FRDs

The outcomes of the statistical analysis showed that survey respondents had a strong preference for feeding FRDs ($\chi^2 = 423.18$, df = 2, p < 0.001). Additionally, a significantly high section of the responders confirmed use of lemon in their diet ($\chi^2 = 171.18$, df = 2, p < 0.001). In addition, it was noted that food items were often prepared and served with lemon for human consumption, before the leftovers were offered to FRDs ($\chi^2 = 22.207$, df = 2, p < 0.001).

On the other hand, participants who had pet dogs avoided adding lemon to their pet dogs' food ($\chi^2 = 294.04$, df = 2, p < 0.001). Participants also mentioned that they incorporate lemon into regular food as a habit, and unintentionally gave it to the FRDs ($\chi^2 = 213.57$, df = 1, p < 2.2e-16). These results provide significant context for understanding participants' feeding habits and regards toward FRDs. They also emphasize the unintentional addition of lemon to FRDs' diets, even if pet owners intentionally avoid adding it to their own dogs' food.

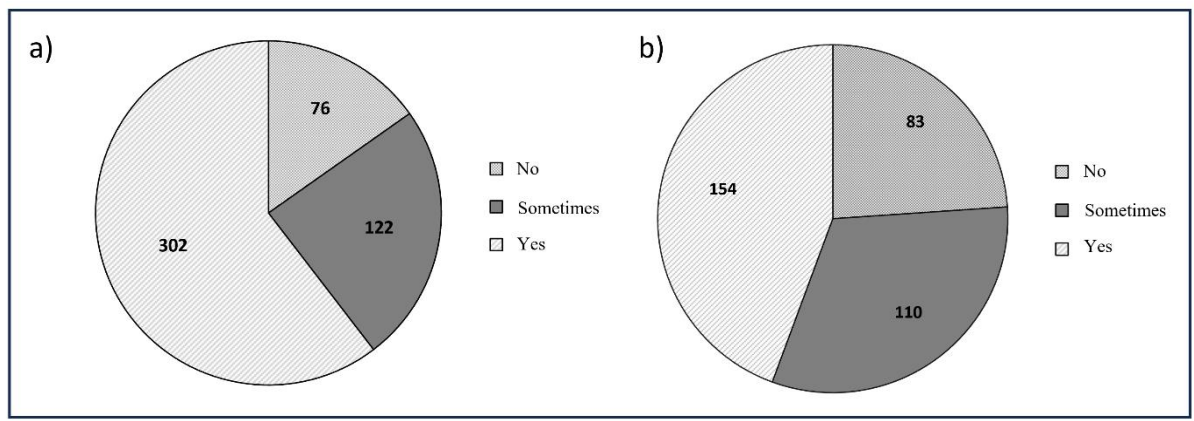

Fig. 2 Pie charts representing peoples' responses. a) People's responses on whether they use lemon in their daily diet b) People's responses on whether they mix lemon parts with food when they feed FRDs.

### B. Investigating food preference and citrus avoidance behaviour in Indian free-ranging dogs: Two-bowl choice test

In the Two Bowl Choice Test, a total of 196 sniffing occurrences were recorded for biscuits, while 165 sniffing occurrences were observed for lemons. Statistical analysis revealed no significant difference between the frequencies of sniffing events for the two food categories ($\chi^2 = 1.167$, df = 1, p-value = 0.28). Additionally, examination of first sniffing events indicated that biscuits were first sniffed 103 times, whereas lemons were first sniffed 93 times, with no significant difference observed between the first sniffing events for biscuits and lemons ($\chi^2 = 0.16337$, df = 1, p-value = 0.686). A notable disparity in eating preferences was evident, as biscuits were consumed 196 times, while lemons were not eaten at all. This discrepancy in eating behaviours resulted in a significant difference in eating preferences between biscuits and lemons ($\chi^2 = 128.01$, df = 1, p < 0.001).

Further, SnS strategy occurred with biscuits (196 occurrences). The dogs exhibited a significant difference between SnS and non-SnS behaviours for biscuits ($\chi^2 = 128.01$, df = 1, p < 0.001). Additionally, post-choice sniffing behaviours were investigated, revealing that after selecting a biscuit, the same was sniffed 93 times, signifying a significant difference between post-choice sniffing and non-sniffing behaviours for biscuits ($\chi^2 = 59.362$, df = 1, p < 0.001). Conversely, after opting for a biscuit, lemons were sniffed 72 times and not sniffed 31 times, depicting a significant difference between post-choice sniffing and non-sniffing behaviours for lemons ($\chi^2 = 7.688$, df = 1, p-value = 0.0055). These findings reveal that the dogs are exploring both available choices before discarding the lemon, suggesting decision-making processes being involved, and not impulsive selection of a food item.

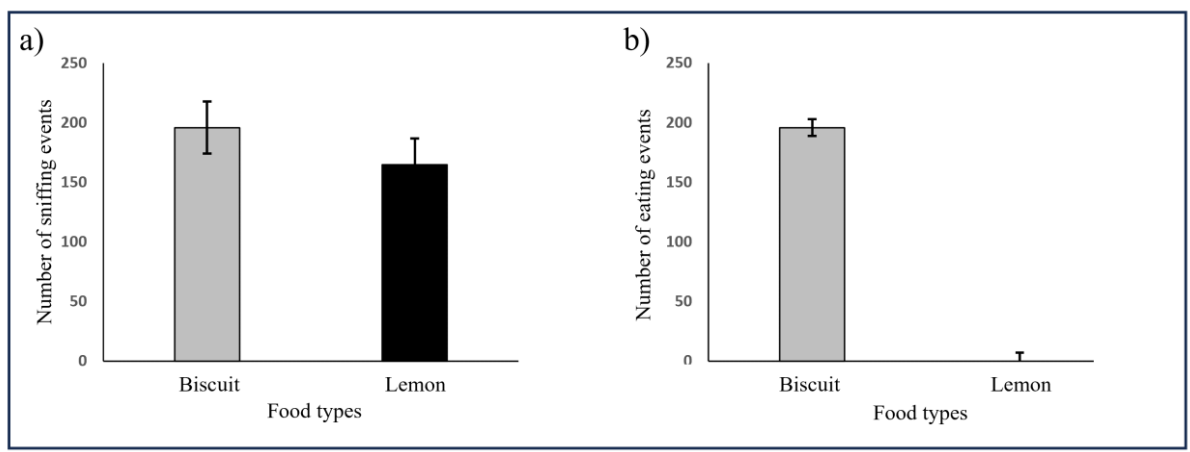

Fig. 3 Bar plots representing the Mean ± S.D. frequency of different behaviours for different food types: a) Frequency of sniffing behaviour for biscuit and lemon b) Frequency of eating behaviour for biscuit and lemon.

### C. Three bowl choice tests

Chicken with lemon pulp (LP), lemon rind (LR), and lemon juice (LJ) were sniffed in initial choice 37, 38, and 36 times, respectively. The chi-square goodness of fit test found no significant difference in the likelihood of first sniffing among the three food items ($\chi^2 = 0.054$, df = 2, p = 0.98). The likelihood of first sniffing events did not change substantially across the given meal types LP & LR, LP & LJ, and LR & LJ (p > 0.05 for all comparisons).

Chicken with LP, LR and LJ were licked in first choice 41, 31, and 18 times, respectively. A significant difference in the likelihood of first licking across all three food items was observed ($\chi^2 = 8.867$, df = 2, p = 0.012). First licking events were marginally significantly different between food types LP and LJ ($\chi^2 = 8.97$, df = 1, p = 0.0027), but not between LP and LR or LR and LJ (p > 0.05). All three food items received similar levels of vigorous licking (37, 38, and 36 times, respectively) ($\chi^2 = 2.641$, df = 2, p = 0.267).

The frequencies of consumption of chicken with LP, LR, and LJ as the first choice, were 42, 28, and 4, respectively, with their likelihood of initial consumption being significantly different ($\chi^2 = 29.946$, df = 2, p < 0.001). Chicken with LJ was consumed lesser than the LP ($\chi^2 = 31.391$,

df = 1, p < 0.001) and LR ($\chi^2$ = 18, df = 1, p < 0.001) ones. No significant difference was found in the first choice of consumption between chicken with LP and LR ($\chi^2$ = 2.8, df = 1, p = 0.094).

The total number of times the food items was selected for consumption showed significant difference among the three food types ($\chi^2$= 31.032, df = 2, p < 0.001) as chicken with LP, LR, and LJ were consumed 64, 46, and 14 times, respectively. Chicken with LJ was significantly less consumed than the LP ($\chi^2$ = 32.051, df = 1, p < 0.001) and LR ($\chi^2$ = 17.067, df = 1, p < 0.001) ones. No significant difference was found in the total number of times chicken with LP and LR were consumed ($\chi^2$ = 2.94, df = 1, p = 0.086).

The occurrence of SnS was recorded 61 times, and was absent 50 times. A chi-square goodness of fit test conducted to assess the likelihood of dogs engaging in SnS versus not following it for chicken with LP revealed no significant difference ($\chi^2$= 0.36576, df = 1, p-value = 0.54). SnS was documented 38 times, contrasting with its absence in 73 instances for chicken with LR ($\chi^2$= 5.0304, df = 1, p-value = 0.025). In the context of chicken with LJ, SnS was observed 13 times, while it was absent in 98 cases ($\chi^2$= 36.362, df = 1, p < 0.001). Thus, the dogs on the whole refrained from using SnS, a strategy that they use to quickly eat a preferred food, when chicken was contaminated with lemon.

The results of the correlation test revealed a moderate positive correlation (p-value < 0.001) and a weak positive correlation (0.189) between eating and sniffing events (p-value < 0.001), as well as a strong positive correlation (0.656) between eating and licking events (p-value < 0.001) (Supplementary table 1). These findings suggest that dogs who liked the food items more also eventually ate more, while the relationships between licking and sniffing and eating and sniffing were progressively weaker but still statistically significant.

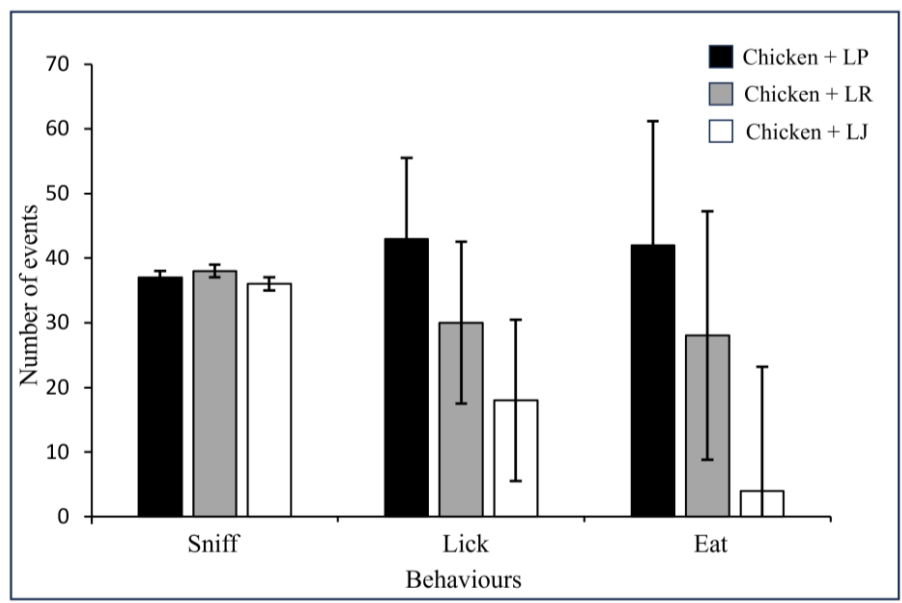

Fig. 4 Bar plot representing the Mean ± S.D. of the frequency of different behaviours observed for different parts of lemon used to contaminate the chicken.

**D. Three-bowl, three concentration choice tests**

After placement of the set-up, dogs approached and sniffed the foods provided in all successful trials. No significant difference emerged in their first sniffing events among the three bowls ($\chi2 = 2.7222$, df = 2, p = 0.256). Dogs showed varying preferences for licking chicken from bowls with different lemon concentrations ($\chi2 = 7.4286$, df = 2, p = 0.024). Specifically, they licked chicken less from the 50% LJ bowl than the 25% LJ one ($\chi2 = 7.3333$, df = 1, p = 0.007). Significant differences emerged in strategizing events before eating, among the three bowls ($\chi2 = 15.242$, df = 2, p < 0.001). Pairwise comparisons revealed significant differences between 50% and 25% LJ bowls ($\chi2 = 13.235$, df = 1, p < 0.001) and between 33.3% and 25% LJ bowls ($\chi2 = 6.3684$, df = 1, p = 0.012). No significant difference was found between 50% and 33.3% LJ bowls ($\chi2 = 1.3913$, df = 1, p = 0.238).

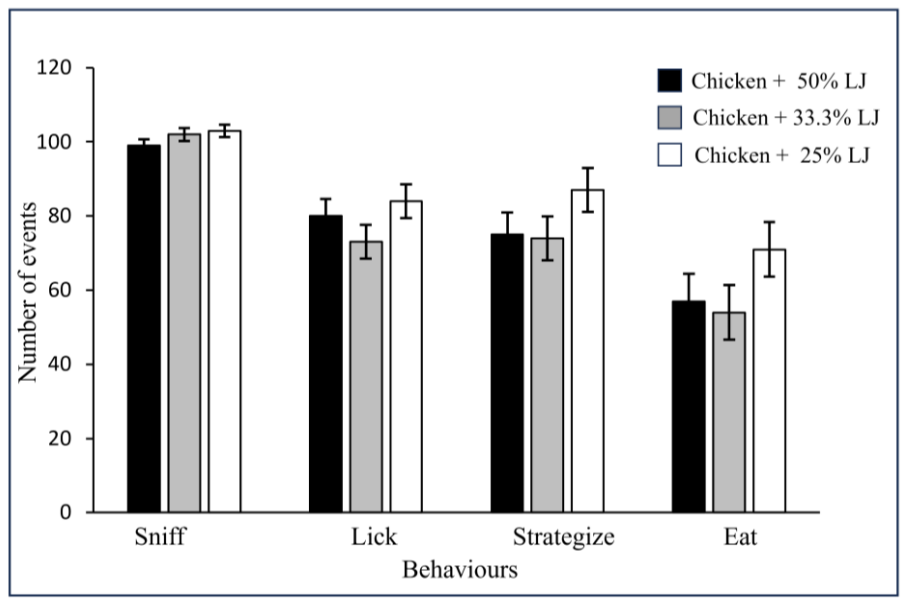

Fig. 5 Bar plot representing the Mean ± S.D. of the frequency of different behaviours observed for different concentrations of LJ used with chicken.

Significant differences were observed in dogs' first eating preferences among the three bowls ($\chi2 = 15.519$, df = 2, p < 0.001). Pairwise comparisons indicated significant differences between first eating from 50% and 25% LJ bowls ($\chi2 = 14$, df = 1, p < 0.001) and between 33.3% and 25% LJ bowls ($\chi2 = 5.5538$, df = 1, p = 0.01844). The time spent by the dogs in the first encounter across three bowls with concentrations of 50%, 33.3%, and 25% were significantly different (Kuskal-Wallis test; H = 10.432, df = 2, p = 0.005). Subsequent post-hoc analysis using Dunn's test identified specific pairwise differences among the concentrations. Significant differences were observed between the 50% and 25% concentrations (p = 0.001) as well as between the 33.3% and 25% concentrations (p = 0.006). These p-values were adjusted using the Bonferroni correction method, with a corrected alpha of 0.0167, to account for multiple comparisons. Overall, these findings suggest significant differences in involvement time among the concentrations (Test power = 0.972). The total engagement time of the dogs across three bowls with concentrations of 50%, 33.3%, and 25% was not statistically different (Kruskal-Wallis test; H = 1.9891, df = 2, p = 0.37).

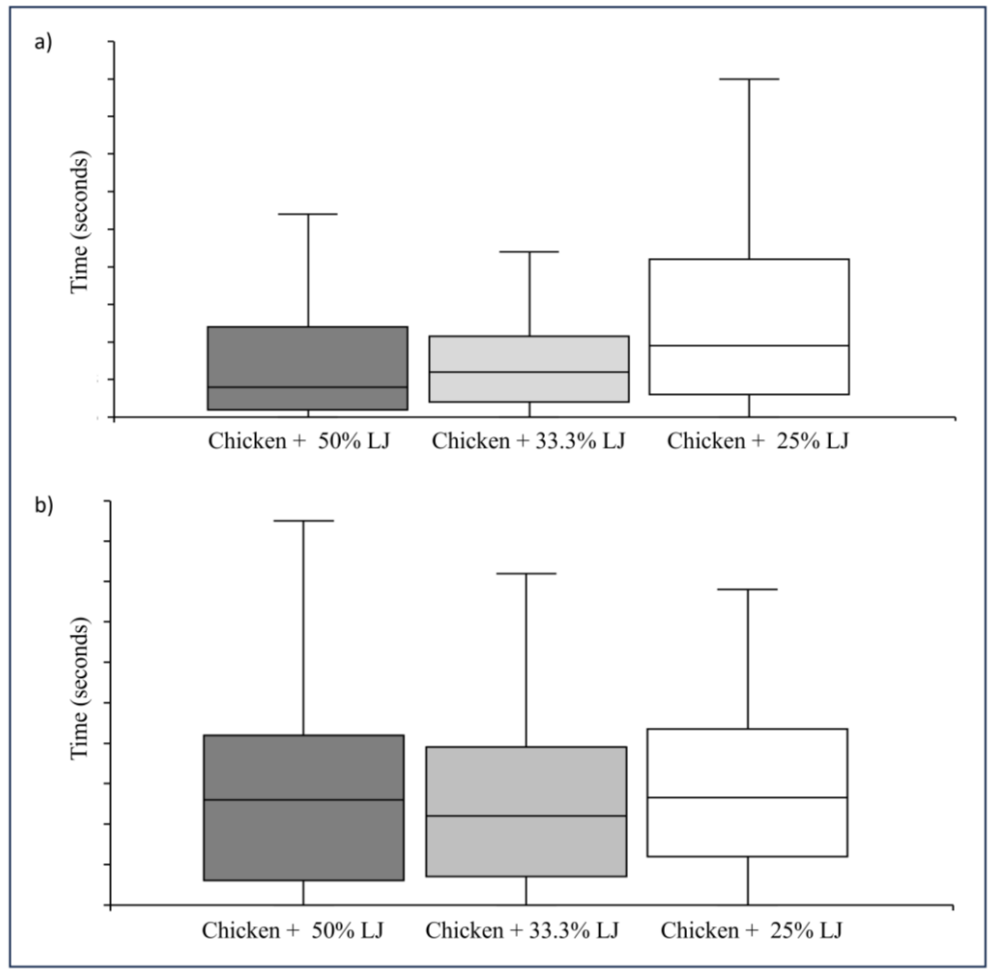

Fig. 6 Box and whiskers plot representing the interaction time for different concentrations of lemon juice used with chicken: a) Time investment in the first interaction with the three different bowls b) Total time investment in interaction with the three different bowls.

**Duration of interaction with the food**

To understand which factors, influence the first interaction time (duration of time spent in the first interaction with the food bowls), we perform a beta regression model with a logit link function.

betareg(formula = First_int_time ~ Conc + Sex + Eat + Latency, data = datasheet)

Higher First Interaction Time (interaction time during the first encounter) was found to be correlated with higher concentration levels when compared to the 25% reference level, indicating a dose-dependent effect. The First Interaction Time was significantly affected by concentrations of 33.3% ($p < 0.01$) and 50% ($p < 0.0001$). Furthermore, a positive coefficient (0.944) and an extremely significant p-value ($p < 1e-13$) indicate a strong association between eating and shorter First Interaction Times.

Conversely, there was a marginally negative correlation with latency, but it was not statistically significant (p > 0.1). Moreover, there was no statistically significant difference in the impact of gender (p > 0.60).

To understand which factors, affect the total interaction time (total time spend to interact with food and bowl), another beta regression analysis was performed with a logit link function.

Betareg (formula = Total_int_time ~ Conc + Sex + Eat + Latency, data = datasheet)

Eating was positively correlated with total interaction time (p < 0.0001), but there was no statistically significant difference between concentration at 33.3% and 50% and the reference level of 25%. It means that eating and total invested time are associated. Neither latency nor sex significantly affected total interaction time (p > 0.6 for both). According to the model's fit indicators (pseudo-R-squared = 0.1934), the variation in total interaction time can be explained to a considerable extent.

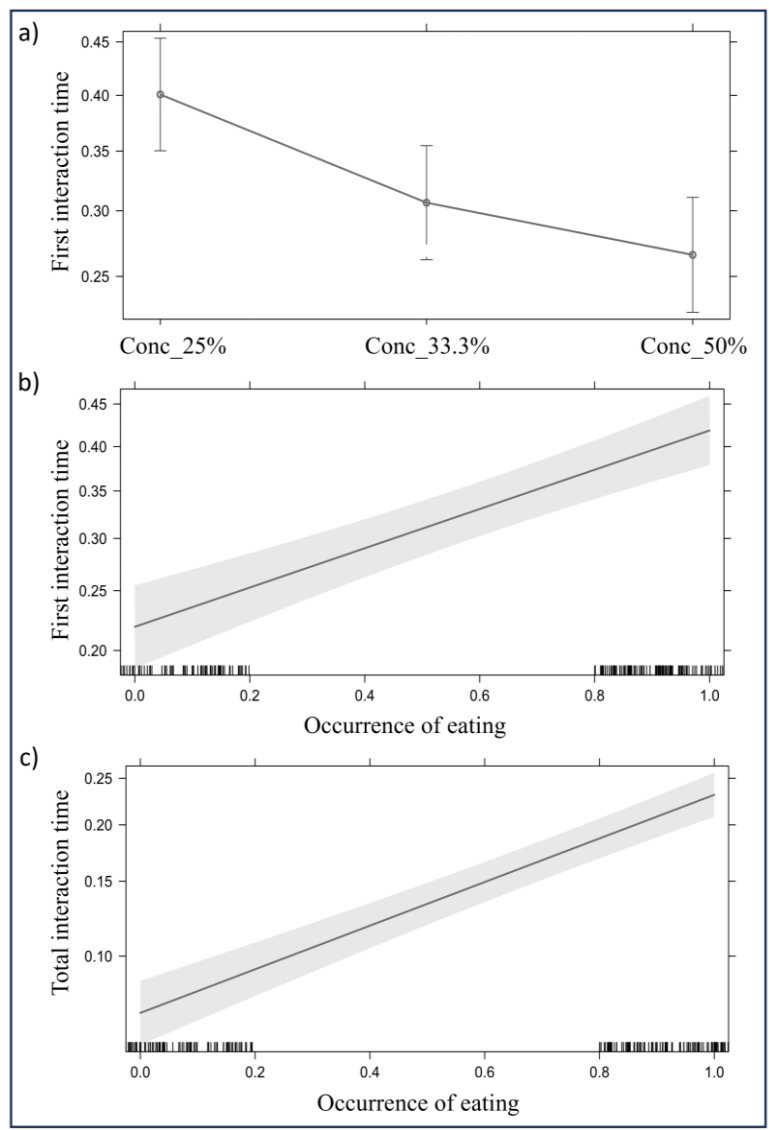

Fig. 7 Statistical modelling results: a) The effect of the concentration of the LJ on the time of first interaction b) The effect of the complete eating of the chicken piece on the time of first interaction c) The effect of the complete eating of the food provided on the time of total interaction.

**Factors Influencing the Likelihood of Eating**

A logistic regression model was used to identify the factors that influence the likelihood of eating the food.

Eat ~ Conc +Sex + First_int_time + Total_int_time + Latency, data = datasheet, family = "binomial"

Based on this model, total time investment was the sole significant predictor that affected the likelihood of eating ($p < 3.4e-11$), and the positive estimate (8.689) shows that the log odds of eating increase as total time investment increases. There were no obvious impacts of latency, sex, or concentration on eating. AIC = 380.59 indicates a moderate level of model fit.

To understand the acidity of the lemon juice solutions, we measured the pH of different concentration of the lemon juice solutions. We found that the more concentrated solutions were more acidic. The average pH values were 2.94, 3.25, and 3.87 for the 50%, 33.3%, and 25% concentrations, respectively.

**Discussion:**

Our study had two parts – (a) the survey of people to understand their perceptions of FRDs and their tendencies to feed the dogs; and (b) the response of FRDs to food contaminated by lemon in various ways. The survey revealed a strong tendency of people to interact with and feed FRDs. This, in association with the reported tradition of preparing and giving food to FRDs with lemon (P. Sarkar et al., 2015; Sen, 2008)and the prevalent use of lemon in the diet of the participants, suggests the possibility of FRDs being exposed to lemon while scavenging among human-generated food waste. However, a crucial finding is that participants who had pet dogs strictly avoided adding lemon to their pet dog's food, which suggests that there may be variation in how people treat FRDs and their own pets.

The two-bowl choice test results provide significant details on the dietary preferences of FRDs. Dogs exhibited a definite preference for non-citrus food, as indicated by the fact that they overwhelmingly preferred biscuits to lemons, even though there was no statistical difference in the frequency of initial sniffing between the two types of biscuits. A particular activity known as Sniff-&-Snatch (SnS), which is typically observed when the FRDs encounter a preferred food item, was only noted when interacting with biscuits. Notably, in earlier choice tests involving biscuits/bread and chicken, SnS was observed for chicken (Bhadra et al., 2016a). This pattern of behaviours indicates a more calculated approach to preferred food items, potentially involving smelling the food first (Cameron et al., 2019). During the test, the dogs engaged in a sampling behaviour, where they would sniff both food items before making a choice. This sampling process is crucial as it allows the dogs to assess the palatability and potential nutritional value of the food items before consumption(Tobie et al., 2015). Dogs have sniffed the uneaten lemons after eating biscuits, suggesting persisting curiosity or maybe hoping for an alternate outcome (Siniscalchi et al., 2011). These behaviours can be interpreted through the lens of Optimal Foraging Theory, which posits that animals maximize their energy intake by selecting more rewarding food items (Pyke, 1984). The preference for biscuits over lemons aligns with this theory, as dogs optimize their foraging efficiency by choosing the more palatable and familiar food over the less palatable one.

When the dogs first sniffed the three different foods i.e. chicken with LP, LJ and LR, they did not show any preference, but they displayed a tendency to lick the chicken with LP more frequently than the one with LJ. In addition, a smaller number of individuals consumed chicken with LJ than chicken with LP or LR. Furthermore, there were differences in the frequency of SnS strategy across all the different food types, suggesting that behavioural reactions differed based on the type of food stimulus that was given. In particular, dogs showed a noticeable inclination to use SnS strategy when provided with chicken with lemon pulp, indicating possible preferences and strategies in their interactions with food. Licking behaviour, particularly the microstructure of licking, is a valuable tool for assessing food preferences and aversions, revealing that while learned hedonic responses can influence behaviours, they are not the sole determinants, and learned changes in consumption patterns may be more resistant to extinction than changes in hedonic responses (Dwyer, 2012). Eating, licking, and sniffing the food items were significantly correlated with one another. In particular, the strong positive correlation suggests that dogs who lick the food are more likely to eat it. Additionally, there is a moderate relationship between licking and sniffing and a noticeable but weaker relationship between eating and sniffing. These findings demonstrate how closely related these behaviours are to one another. Eating and licking showed the strongest correlation, which suggests that once a dog commits to investigate an item of food closely, it is most likely to eventually eat it. This commitment is decided by the first response towards the food, i.e., sniffing. When presented with three options, namely, a piece of chicken with lemon pulp, lemon rind, and lemon juice, the FRDs ate the chicken from both LP and LR more often than the one with LJ. This is most likely because, the lemon juice completely covers the chicken piece and perhaps even seeps into it, and is more difficult to separate the chicken from lemon juice than the lemon pulp or rind.

In the three-bowl choice test with various concentrations of lemon juice, dogs did not exhibit a preference when first sniffing around the bowls. But when they licked first, they showed a preference for the 25% concentration over the 50% concentration. Thus, they are most likely capable of detecting the higher concentration of lemon juice through sniffing, and are thus demonstrating a preference for the lesser contamination available, thereby choosing the bowl with the most dilute lemon juice. The dogs spent more time interacting with the bowls with the 50% and 33.3% concentrations of lemon juice during initial exploration than they did with the 25% concentration, indicating that higher concentrations of lemon juice required more investigation before approaching. It's interesting to note that the overall amount of time spent interacting with the bowls did not significantly change across concentrations, suggesting that the entire engagement time was consistent after a decision was made, perhaps based on the first encounter with the bowls.

The beta regression analysis showed that increased lemon juice concentration led to longer initial interactions with the bowls, indicating a possibility of an initial dislike for stronger lemon flavours. Eating success predicted more extended interaction times, suggesting that once dogs decided to eat, they spent more time with the bowls. Neither gender nor latency significantly affected interaction time and eating success, with total interaction time predicting the likelihood of eating. This suggests a generalist foraging strategy, where dogs adjust their food selection based on availability and palatability (R. Sarkar et al., 2019).

Animals often avoid sour foods because they can be unpalatable and potentially harmful (Mennella et al., 2003). Sourness can signal the presence of unripe or spoiled food, which might contain toxins or lack nutritional value. From an evolutionary perspective, the ability to discern and avoid sour-tasting food has significant fitness benefits (Frank et al., 2022). By selecting palatable over unpalatable food, animals increase their chances of ingesting safe, nutrient-rich food that supports their growth, health, and reproduction (Jacobs et al., 2009).

The sense of taste in nonhuman primates, for instance, has been studied anatomically (Rolls, 1989), behaviourally (Laska, 1996, 1999) and electro physiologically (Plata-Salaman et al., 1995). These studies reveal how primates use gustatory cues to assess the palatability of potential food items, particularly focusing on the balance of sweetness and sourness. This balance varies with the ripeness of fruits and indicates their nutritional value (Chapman, 1987; Ross, 1992).

Our study sheds light on the food habits and foraging habits of Indian FRDs, which can help shape future studies on animal behaviour and push toward responsible human-animal interactions and population management. We explored how Indian FRDs acquired food, paying particular attention to how they behaved to citrus-based food. The outcomes highlighted interesting trends. First, although displaying equal interest in both citrus and non-citrus food during initial sampling, FRDs clearly preferred non-citrus choices, especially biscuits over lemons. This implies an obvious preference for familiar tastes. "Sniff-&-Snatch" was only noticed when interacting with biscuits, suggesting a purposeful approach to desired food items (R. Sarkar et al., 2019). They showed a more complex reaction to the various lemon components—pulp, rind, and juice—when these were given with chicken. Based on the particular lemon component, the dogs modified their strategies, indicating an intricate sense of both texture and flavour. During the three-bowl trial with lemon juice, dogs consistently showed an aversion for higher citrus flavours, favouring bowls with less lemon juice in them. This demonstrates how sensitive they are to flavour intensity, while simply sniffing the food (Kokocińska-Kusiak et al., 2021). Overall, these results highlight how adaptive FRDs' foraging behaviour is, providing insights into how taste preferences affect their selection of food. In order to improve our knowledge of FRDs' behaviour and develop methods for responsible human-animal interactions, future research should also examine the possibility of conditioning taste preferences and carry out comparable studies in various locations to assess cultural influences.

**Ethical Statement**

The study design did not violate the Animal Ethics regulations of the Government of India (Prevention of Cruelty to Animals Act 1960, Amendment 1982). The protocol for the experiment was approved by the IISER Kolkata Animal Ethics Committee.

**Conflict of interest statement**

All the authors have read and agree with this version of the manuscript. The authors declare no conflict of interest.

**Authors' contributions**

The study was performed by TSP started from his MS Thesis with the help of SN and RS. With the help of RS and guidance of AB, TSP designed the experiment. AB supervised the work and gave valuable inputs in writing the paper. Analysis was performed by TSP with the assistance of RS.

**Acknowledgements**

The authors thank Aesha Lahiri, Hindolii Gope, Epil Mandi and Anamitra Roy for their help in conducting parts of the fieldwork. Special thanks to Ashim Kumar Basumatary for his assistance during the study. We also thank Dr. Udipta Chakraborti and Dr. Rubina Mondal for their guidance and valuable suggestions. We are grateful to Sagarika Biswas for creating the illustrations.

**Funding**

TSP was supported by University Grants Commission (UGC). The study is also supported by IISER Kolkata and the Janaki Ammal – National Women Bioscientist Award, Department of Biotechnology. SN was supported by INSPIRE Fellowship, Department of Science and Technology. RS was supported by IISER Kolkata Institute Fellowship.

**Supplementary data**

Will be made available on publication of the manuscript after peer review.

----------

**Supplementary Materials:**

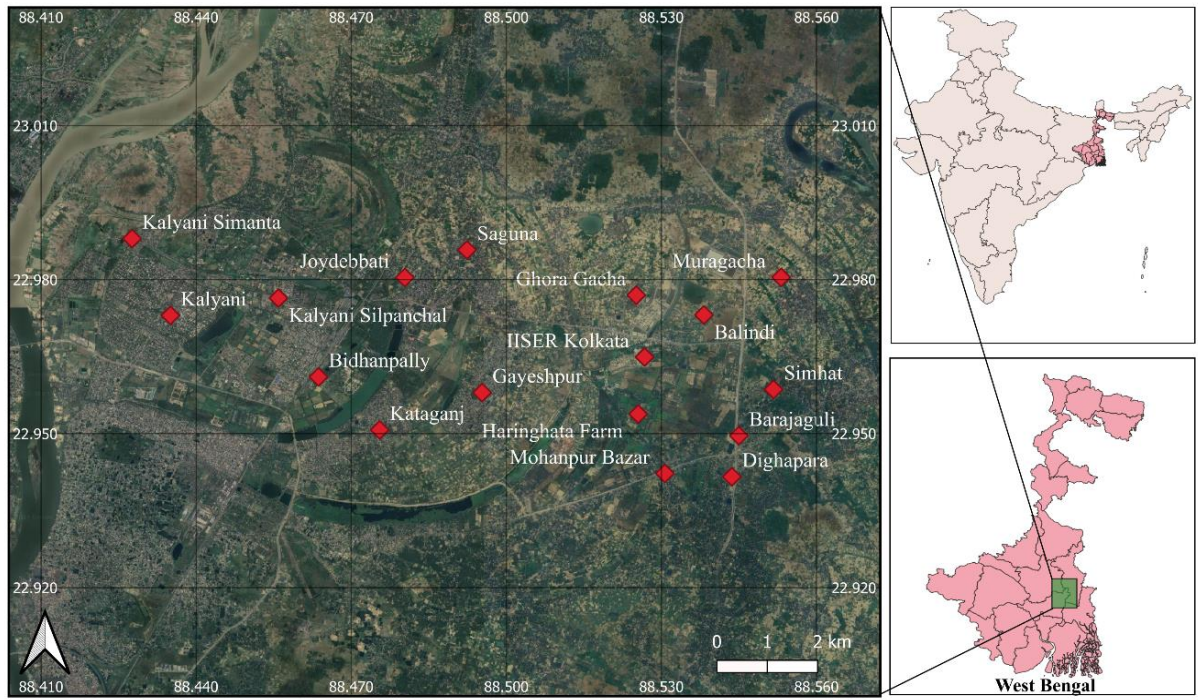

**Fig 1. Study sites**

**Table 1.**
**Correlation values for Three bowl choice test (Among LP, LR & LJ)**

| Correlation Test | t value | df | p-value | 95% Confidence Interval | Correlation coefficient (cor) |
|---|---|---|---|---|---|
| Sniffing and Licking Events | 5.0697 | 331 | 6.644e-07 | 0.1657346 to 0.3653642 | 0.268429 |
| Sniffing and Eating Events | 3.5108 | 331 | 0.0005087 | 0.08370278 to 0.29102342 | 0.189474 |
| Licking and Eating Events | 15.812 | 331 | < 2.2e-16 | 0.5901067 to 0.7131745 | 0.6559789 |

**Table 2. Chi Square test**

| Sl no. | Test | Topic | $\chi^2$ | df | P value |
|---|---|---|---|---|---|
| 1. | Perceptions and practices concerning food provisions for free-ranging and pet dogs | Feeding and interacting with FRDs | 423.18 | 2 | 2.2E-16 |
| | | Lemon consumption among participants | 171.18 | 2 | 2.2E-16 |
| | | Foods often prepared and served with lemon | 22.21 | 2 | 1.506E-5 |
| | | Participants avoided adding lemons to the pet dogs' food | 294.04 | 2 | 2.2e-16 |
| | | Unintentional food mixing with lemon | 213.57 | 1 | 2.2e-16 |
| 2. | Food preference study between two different food items | Total sniffing events | 1.167 | 1 | 0.28 |
| | | First sniffing events | 0.163 | 1 | 0.686 |
| | | Eating events | 128.01 | 1 | 2.2e-16 |
| | | SnS vs not SnS strategy for Biscuit | 128.01 | 1 | 1.312e-14 |
| | | SnS vs not SnS strategy for Lemon | 128.01 | 1 | 2.2e-16 |
| | | Sniffing event of lemon after first choice of biscuit | 7.69 | 1 | 0.005 |
| | | Sniffing event of biscuit after first choice | 59.362 | 1 | 1.312e-14 |
| 3. | Preference test of different parts of citrus fruit | Total sniffing events | 0.053 | 2 | 0.974 |
| | | First sniffing events | 0.024 | 2 | 0.987 |
| | | Total licking evets | 3.36 | 2 | 0.186 |
| | | First licking events | 7.10 | 2 | 0.029 |
| | | First licking events (LP vs LS) | 0.45 | 1 | 0.504 |
| | | First licking events (LP vs LJ) | 3.88 | 1 | 0.048 |
| | | First licking events (LJ vs LS) | 1.26 | 1 | 0.262 |
| | | Total vigorous licking | 2.641 | 2 | 0.267 |

| | | Total eating events | 24.308 | 2 | 5.27e-06 |
| --- | --- | --- | --- | --- | --- |
| | | Total eating events (LP vs LS) | 12.55 | 1 | 0.0004 |
| | | Total eating events (LP vs LJ) | 23.09 | 1 | 1.542e-06 |
| | | Total eating events (LJ vs LS) | 1.661 | 1 | 0.19 |
| | | First eating events | 21.587 | 2 | 2.053e-05 |
| | | First eating events (LP vs LS) | 8.8 | 1 | 0.003 |
| | | First eating events (LP vs LJ) | 16.985 | 1 | 3.768e-05 |
| | | First eating events (LJ vs LS) | 1.039 | 1 | 0.308 |
| | | Proportion of SnS for LP | 0.366 | 1 | 0.545 |
| | | Proportion of SnS for LS | 5.030 | 1 | 0.025 |
| | | Proportion of SnS for LJ | 36.362 | 1 | 1638e-09 |
| 4. | Preference test for three distinct concentrations (50%, 33.3% and 25%) of lemon juice with chicken | First sniffing events | 2.272 | 2 | 0.256 |
| | | First licking events (50% vs 33.3%) | 1.852 | 1 | 0.174 |
| | | First licking events (25% vs 33.3%) | 1.895 | 1 | 0.169 |
| | | First licking events (25% vs 50%) | 7.333 | 1 | 0.007 |
| | | Strategy making before eating | 15.242 | 2 | 0.0005 |
| | | Strategy making before eating (50% vs 33.3%) | 1.391 | 1 | 0.2382 |
| | | Strategy making before eating (25% vs 33.3%) | 6.368 | 1 | 0.0116 |
| | | Strategy making before eating (25% vs 50%) | 13.235 | 1 | 0.0003 |
| | | First eating events | 15.519 | 2 | 0.0004 |
| | | First eating events (50% vs 33.3%) | 2.1892 | 1 | 0.139 |
| | | First eating events (25% vs 33.3%) | 5.554 | 1 | 0.018 |
| | | First eating events (25% vs 50%) | 14 | 1 | 0.0002 |

**Statistical modelling results 1.**
1. model1 <- betareg(First_int_time ~ Conc + Sex + Eat + Latency, data = rrr

**Result:**

Standardized weighted residuals 2:
   Min    1Q  Median    3Q    Max
-6.1932 -0.3076  0.2185  0.6275  7.1453

Coefficients (mean model with logit link):
                   Estimate Std. Error  z value  Pr(>|z|)
(Intercept)       -0.818783  0.165740  -4.940  7.81e-07 ***
ConcConc_33.3%    -0.414244  0.154904  -2.674  0.00749 **
ConcConc_50%      -0.614011  0.154851  -3.965  7.33e-05 ***
SexM              -0.066636  0.127501  -0.523  0.60123
Eat                0.944470  0.129795   7.277  3.42e-13 ***
Latency           -0.009684  0.006371  -1.520  0.12850

Phi coefficients (precision model with identity link):
      Estimate Std. Error z value Pr(>|z|)
(phi)   1.808    0.128   14.13   <2e-16 ***
---
Signif. codes:  0 '***' 0.001 '**' 0.01 '*' 0.05 '.' 0.1 ' ' 1

Type of estimator: ML (maximum likelihood)
Log-likelihood:   151 on 7 Df
Pseudo R-squared: 0.1064
Number of iterations: 16 (BFGS) + 3 (Fisher scoring)

**2.model4 <- betareg(Total_int_time ~ Conc + Sex + Eat + Latency, data = rrr)**

**Result:**

Standardized weighted residuals 2:
    Min    1Q  Median    3Q    Max

-9.2112 -0.2645  0.1920  0.6795  3.4939

Coefficients (mean model with logit link):
                  Estimate  Std. Error  z value  Pr(>|z|)
(Intercept)      -2.635261   0.145161  -18.154   <2e-16 ***
ConcMod_BTF_33    0.086720   0.126996    0.683    0.495
ConcMod_CTF_25    0.146502   0.126139    1.161    0.245
SexM              0.048314   0.105021    0.460    0.645
Et_1_0            1.347102   0.111803   12.049   <2e-16 ***
Latency          -0.002668   0.005317   -0.502    0.616

Phi coefficients (precision model with identity link):
     Estimate Std. Error z value Pr(>|z|)
(phi)  5.0306   0.4205    11.96  <2e-16 ***
---
Signif. codes:  0 '***' 0.001 '**' 0.01 '*' 0.05 '.' 0.1 ' ' 1

Type of estimator: ML (maximum likelihood)
Log-likelihood: 398.4 on 7 Df
Pseudo R-squared: 0.1934
Number of iterations: 26 (BFGS) + 1 (Fisher scoring)

**3.model7 <- glm(Eat ~ Conc +Sex + First_int_time + Total_int_time + Latency, data = datasheet, family = "binomial")**
**Result:**

Deviance Residuals:
   Min     1Q   Median     3Q     Max
-2.9478 -0.8819  0.3987  0.9367  1.7304

Coefficients:
                 Estimate  Std. Error  z value  Pr(>|z|)
(Intercept)     -0.657938   0.375654   -1.751   0.0799 .
ConcConc_33.3%  -0.544176   0.314939   -1.728   0.0840 .
ConcConc_50%    -0.465219   0.318505   -1.461   0.1441
SexM            -0.125798   0.258195   -0.487   0.6261
First_int_time   0.319620   0.601650    0.531   0.5953
Total_int_time   8.689758   1.311091    6.628   3.41e-11 ***
Latency         -0.004401   0.013010   -0.338   0.7352
---
Signif. codes:  0 '***' 0.001 '**' 0.01 '*' 0.05 '.' 0.1 ' ' 1

(Dispersion parameter for binomial family taken to be 1)

    Null deviance: 444.69  on 323  degrees of freedom
Residual deviance: 366.59  on 317  degrees of freedom
AIC: 380.59

Number of Fisher Scoring iterations: 5

**Information 1**

## Survey Questionnaire (English) given below

Perception of Food Provisions for Free-Ranging Dogs and Pet Dogs: The Phenomenon of Lemon Inclusion

1. Do you agree to participate the survey?
    - Yes
    - No

2. Sex
    - Male
    - Female
    - Don't want to say
    - Others

3. Age (Above 18yrs) _________________

4. Location (In short) _________________

5. Profession _____________________

6. Do you have any pet dog?
    - Yes
    - No
    - Prefer not to say

7. Specify breed, _______________

8. What type of food do you eat on daily basis?
    - Veg
    - Non-veg
    - Prefer not to say
    - Both

9. What do you feed your pet dog?
    - Veg
    - Non-veg
    - Both

10. Do you like Free-ranging dogs?
    - Yes
    - No
    - Neutral

11. Do you feed Free-ranging dogs?
    - Yes

- No
- Sometimes
- Can't say

12. If you feed FRD's, then you feed which age group of FRD?
    - Pups
    - Juvenile
    - Adult
    - Mixed

13. Which type of foods do you provide to Free-ranging dogs?
    - Veg
    - Non-veg
    - Mixed
    - Others

14. Do you use lemon in your daily diet?
    - Yes
    - No
    - Sometimes

15. Do you add lemon to the food given to your pet dog?
    - Yes
    - No
    - Sometimes

16. Have you given lemon mixed with food to free-ranging dogs?
    - Yes
    - No
    - Sometimes
    - Can't say

17. If yes, have you given food mixed with lemon to the FRD's intentionally or unintentionally
    - Intentionally
    - unintentionally
    - can't say

18. Have you ever seen that dog's eating lemon or lemon shell?
    - Yes
    - No
    - Sometimes
    - Can't say

19. Do you think dogs like the taste of lemon or lemon flavour?
    - Yes
    - No
    - May be

- Can't say
- 

..........